\documentstyle[11pt,newpasp,twoside,epsf]{article}
\def\edcomment#1{\iffalse\marginpar{\raggedright\sl#1\/}\else\relax\fi}
\reversemarginpar

\begin{document} \title{The spectroscopic orbit and other parameters of AE Arae} 


\author{J. Miko{\l}ajewska$^1$,  C. Quiroga$^{2,4}$,  E. Brandi$^{2,3}$, \\ 
L. G. Garc\'{\i}a$^2$, O. E. Ferrer$^{2,4}$ \& K. Belczy{\'n}ski$^5$}

\affil{$^1$Copernicus Astronomical Center, Warsaw, Poland \\ 
$^2$Facultad de Ciencias Astron\'omicas y Geof\'{\i}sicas, Universidad Nacional de La 
Plata, Argentine \\ 
$^3$Comisi\'on de Investigaciones Cient\'{\i}ficas de la Provincia de Buenos Aires (CIC), 
Argentine \\ 
$^4$Consejo Nacional de Investigaciones Cient\'{\i}ficas y T\'ecnicas 
(CONICET), Argentine\\ 
$^5$Department of Physics and Astronomy, Northwestern University, Evanston, USA}

\begin{abstract} We present an analysis of optical and near infrared spectra of the 
symbiotic system AE Ara, composed of an M\,5.5\,III giant similar to the Galactic Bulge M 
giants and a hot luminous companion. In particular,  we have determined for the first time 
spectroscopic orbits  based of the radial velocity curves for the red and the hot 
component from the M-giant absorption lines and from the wings of  He II $\lambda$4686 
emission profiles, respectively. We have also studied spectral  changes and photometric 
variations in function of both  the orbital phase and  activity. The 
resulting physical parameters of the  binary components and the nature of the hot 
component activity are briefly discussed. \end{abstract}

\section{Introduction}

The aim of this work was to determine periods and spectroscopic orbits of symbiotics, for 
which observational data already exists but have yet not been analyzed. High resolution 
spectra for AE Ara were collected with the 2.15 m telescope of CASLEO at San Juan, 
Argentina, during the period 1990$-$2001. We have also collected all published optical 
spectroscopic data (see Miko{\l}ajewska et al. 2002 for references) as well as visual 
photometry from the Variable Star Section Circulars of The Royal Astronomical Society of 
New Zealand (RASNZ; Fig. 1).

\begin{figure}
\plotone{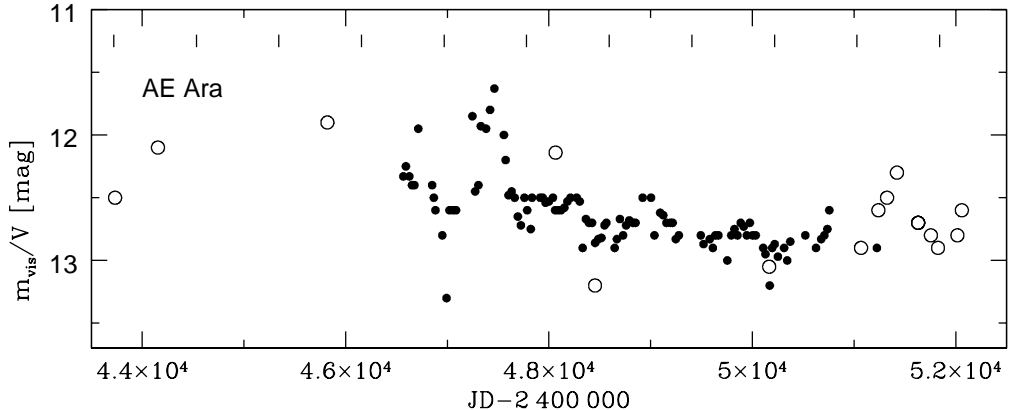} 
\caption{Visual light curve of \mbox{AE Ara} in 1978--98. Dots: visual observations from 
RASNZ, open 
circles: $V$ mags calculated from our spectra, FES counts and literature. Bars mark 
times of photometric minima given by our ephemeris (see text).} 
\end{figure}

We have measured the radial velocities of the cool component from the M-type absorption 
lines. The individual radial velocities were obtained 
by gaussian fitting of the line profiles an a mean value was calculated for each spectrum. 
In addition, we have determined the radial velocities of the broad emission wings of 
H$\alpha$, H$\beta$ and HeII$\lambda$4686 which are formed in the inner region of the 
accretion disk or in an extended envelope near the hot component (see. Quiroga et al. 
2002; Miko{\l}ajewska et al. 2002 for details of the method). 

Optical emission line fluxes were also derived,  either by integrating the line profile or 
by fitting Gaussian profiles, as well as the [TiO]$_1$ and [TiO]$_2$ indices as defined by 
Kenyon \& Fernandez-Castro (1987).\\

\section{Orbital period and spectroscopic orbits}

We have analyzed the RASNZ visual photometry using the period-search method described by 
Schwarzenberg-Czerny (1997). We have found a significant frequency 
peak ($\sim 3 \sigma$) at $\sim 0.00125\ \mbox{day}^{-1}$ which represents light changes 
with amplitude of about 0.4 mag and period of about 800 days.

The radial velocities , emission line fluxes, and [TiO]$_1$ indices (Fig. 2), all show the 
same periodicity, which we attribute to the orbital period. In all cases, the most regular 
changes have been obtained with 812-day period, giving the ephemeris  \begin{equation} 
\mbox{Min} = \mbox{JD}\,2450217\,(\pm 3) + 812\,(\pm 2)\, \times \mbox{E}.  \end{equation} 

The broad emission line wings of  HeII\,4686  show the highest amplitude and a mean 
velocity similar to the red giant systemic velocity. They are clearly in antiphase with 
the M-giant curve which suggest that they are formed in a same region very near to the hot 
component (Fig. 2). Our best orbital solution for the M giant is consistent with a 
circular orbit and the time of the spectroscopic conjunction coincides with the 
photometric minimum. The orbital solutions in Table 1 and Fig. 2 are obtained assuming 
$e=0$ and the spectroscopic conjunction given by Eq.(1). 
\begin{table} 
\caption{Orbital solutions for AE Ara} 
\begin{tabular}{l c c c  l } \hline 
Component & $\gamma\, [\mathrm{km\,s^{-1}}]$ & $K\, [\mathrm{km\,s^{-1}}]$& $f(M)\, 
[M_{\odot}]$&$A \sin i$[AU]\cr 
\hline 
M abs &$- 15.7\pm0.3$&$5.4\pm0.3$ & 
$0.0133\pm0.0021$ &$0.40\pm0.02$ \cr 
Wings (HeII\,4686)& $6.6\pm4.6$ & $23.7\pm6.5$ &$1.12\pm0.57$ &$1.76\pm0.38$\cr Wings 
(H$\beta$)& $-8.9\pm1.5$ & $8.3\pm2.0$ & $0.048\pm0.062$ &$0.62\pm0.12$\cr \hline\cr 
\end{tabular} \end{table} 

Combining the semi-amplitudes of the M giant and the HeII 
emission wing component from Table 1 gives a mass ratio $q=M_{\rm g}/M_{\rm h}=4.4\pm 
1.5$,  the component masses of $M_{\rm g} \sin^3 i \sim  1.7\, \rm M_{\odot}$ and $M_{\rm 
h} \sin^3 i \sim 0.4\, \rm M_{\odot}$, and the binary separation $a \sin i \sim 2 \rm AU.$ 
\\ 
\begin{figure} \plotone{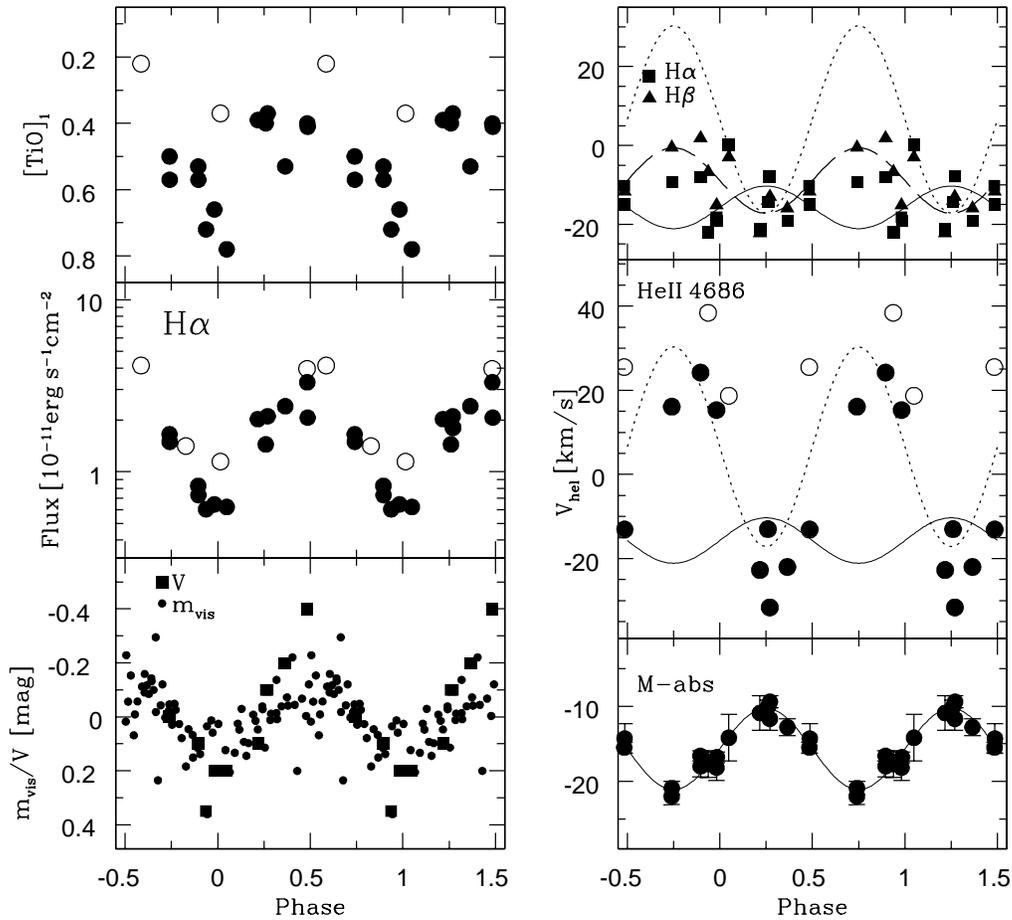} 
\caption{(left) Phase plot of quiescent visual magnitudes, H$\alpha$ emission line fluxes, 
and [TiO]$_1$ indices. (right) Phased radial velocity data and circular orbital solution 
from Table 1. The solid line repeats the orbit of the M giant and the dotted line -- the 
HeII emission wing solution, respectively. Closed symbols refer to quiescent state, open 
circles refer to outburst phase.} \end{figure}

\section{The hot component and its activity}

The visual light curve of \mbox{AE Ara} (Fig. 1)
display a very deep minimum near MJD 46950 ($\phi \sim 0$ according to
our ephemeris)
and two bright maxima around MJD 46700 ($\phi \sim 0.7$)
and MJD 47300-400 ($\phi \sim 0.5-0.6$).
The fact that the minimum is practically as deep as usually strongly suggests
that the hot component is responsible for this brightening.

The optical brightening is accompanied by an increase in emission line fluxes 
(Fig. 2) and broadening of the emission line wings.  Near 
the bright visual maximum around MJD 46700, however,  the HeII\,4686 emission line flux 
was suppressed indicating a decrease of the hot component temperature, possibly because of 
increasing optical depth in the hot component wind. 

The observed changes in the emission line profiles
are typical of symbiotic stars at outbursts,
especially symbiotic novae,
and suggest enhancement of a wind from the hot component surface.
Such wind seems to be permanently present in \mbox{AE Ara}
as suggested by a presence of broad wings in HeII\,4686
emission line, as well as a broad CIV\,5802,5812
emission lines. The CIV lines are typical for 
early type Wolf-Rayet (W-R) WC stars
and subluminous W-R stars found in some planetary nebula,
and they are rarely observed in symbiotic systems. 
For example, they were detected during late outburst phase in the symbiotic
nova AG Peg, and at some phase of the 
recent outburst of \mbox{AS 338}.
In \mbox{AE Ara}, the CIV lines
are almost as strong as HeII\,4686, and their full width
is of about 600 km\,s$^{-1}$, similar to the FW of the HeII line.
Both the CIV and HeII probably arise from the hot component 
wind. The presence of such a wind is also suggested by the large
contribution of the hot component to the optical continuum.
In fact, the $UBV$ colours of the hot component
(derived from the $UBV$ data corrected for contribution
from the M\,5.5 III star) are very similar to those of the hottest
W-R/WC stars, which strongly suggests presence of an extended
expanding atmosphere.
We estimate the mass loss rate
$\dot{M}_{\rm h} \sim {\rm a\,few}\, \times
10^{-8} - 10^{-7} {\rm M_{\odot}\,yr^{-1}}$
from the intensities of HeII\,1640
and HeII\,4686 emission lines.

The presence of an extended photosphere and moderate velocity wind, 
makes it difficult to estimate properly the effective
temperature and luminosity from the available data.
Although, a star with $T_{\rm h} \sim 70\,000 - 80\,000$ K
and $L_{\rm h} \sim 1000 - 2000 \rm L_{\odot}$,
accounts for our observations, both better spectroscopic
observations especially in the ultraviolet as well as
detailed models will be necessary to get accurate
and reliable physical parameters of the hot component.

\acknowledgments 

This research was partly founded by KBN Research Grant No. 5\,P03D\,019\,20.

\end{document}